\documentclass{PoS}

\usepackage{epsfig}
\usepackage{graphicx}
\usepackage{rotating}
\usepackage{amsmath}

\title{Pion and muon yields from a 10\,GeV proton beam on various targets}

\ShortTitle{Pion and muon yields}

\author{\speaker{JOHN BACK}%
         \thanks{This work is supported by the Science and Technology Facilities Council
(United Kingdom).}\\
        University of Warwick\\
        E-mail: \email{j.j.back@warwick.ac.uk}}

\abstract
{We present results from MARS Monte Carlo simulations of the accepted pion and muon
yields when a 10\,GeV parabolic proton beam interacts with a solid or powdered 
(50\% density) tungsten target for a Neutrino Factory. We compare these yields with 
the mercury jet target configuration and find that they all
give similar rates of pion production. Also provided are estimates of the amount
of pion re-absorption inside the target.}

\FullConference{10th International Workshop on Neutrino Factories, Super beams and Beta beams\\
                 June 30 - July 5 2008\\
                 Valencia, Spain}
\begin{document}

\section{Introduction}

The purpose of the target system in a Neutrino Factory is to produce a maximum
number of pions from an intense beam of protons. The pions are captured
using a strong ($\sim$20\,T) solenoidal magnetic field and travel into a channel 
where the decay muons are bunched, cooled and accelerated to the required energies 
before decaying to neutrinos in the storage ring.

Recent experimental work~\cite{RobsPaper} has shown that solid tungsten
can withstand the intense 4\,MW beam power that will be required for
a Neutrino Factory. It may also be possible to use a powdered form of tungsten (with
an effective density not more than 50\%) as the target material.
We use the MARS Monte Carlo simulation package~\cite{MARS} to investigate what pion 
and muon yields can be achieved for a solid or powdered tungsten target and compare 
them with those from the mercury jet scenario. We also provide estimates of pion 
re-absorption inside the target.

\section{Target geometry}

\begin{figure}
\begin{center}
\begin{minipage}[hbt]{0.5\textwidth}
\epsfig{file=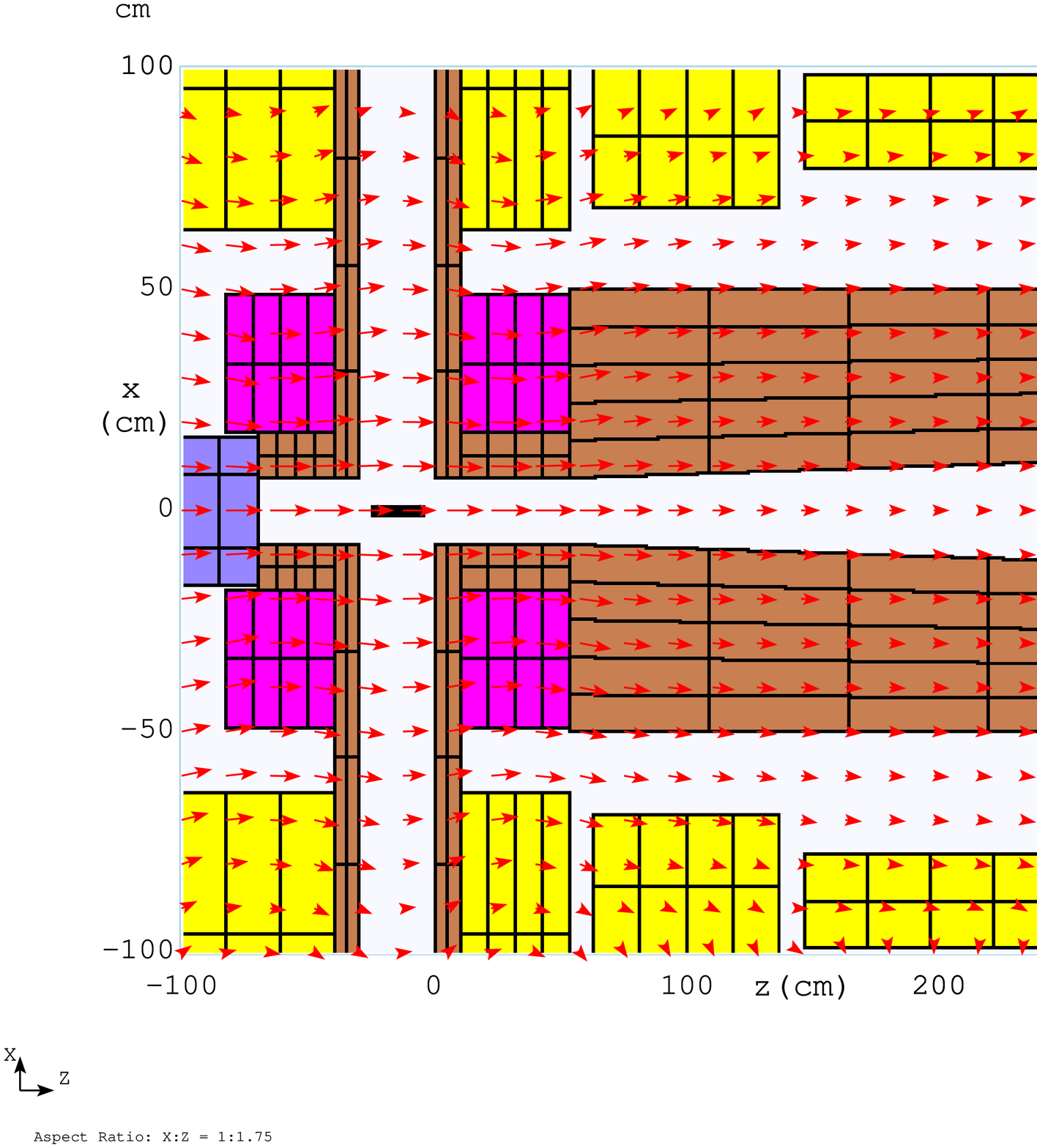, clip=, width=0.8\textwidth}
\end{minipage}\hfill
\begin{minipage}[hbt]{0.5\textwidth}
\epsfig{file=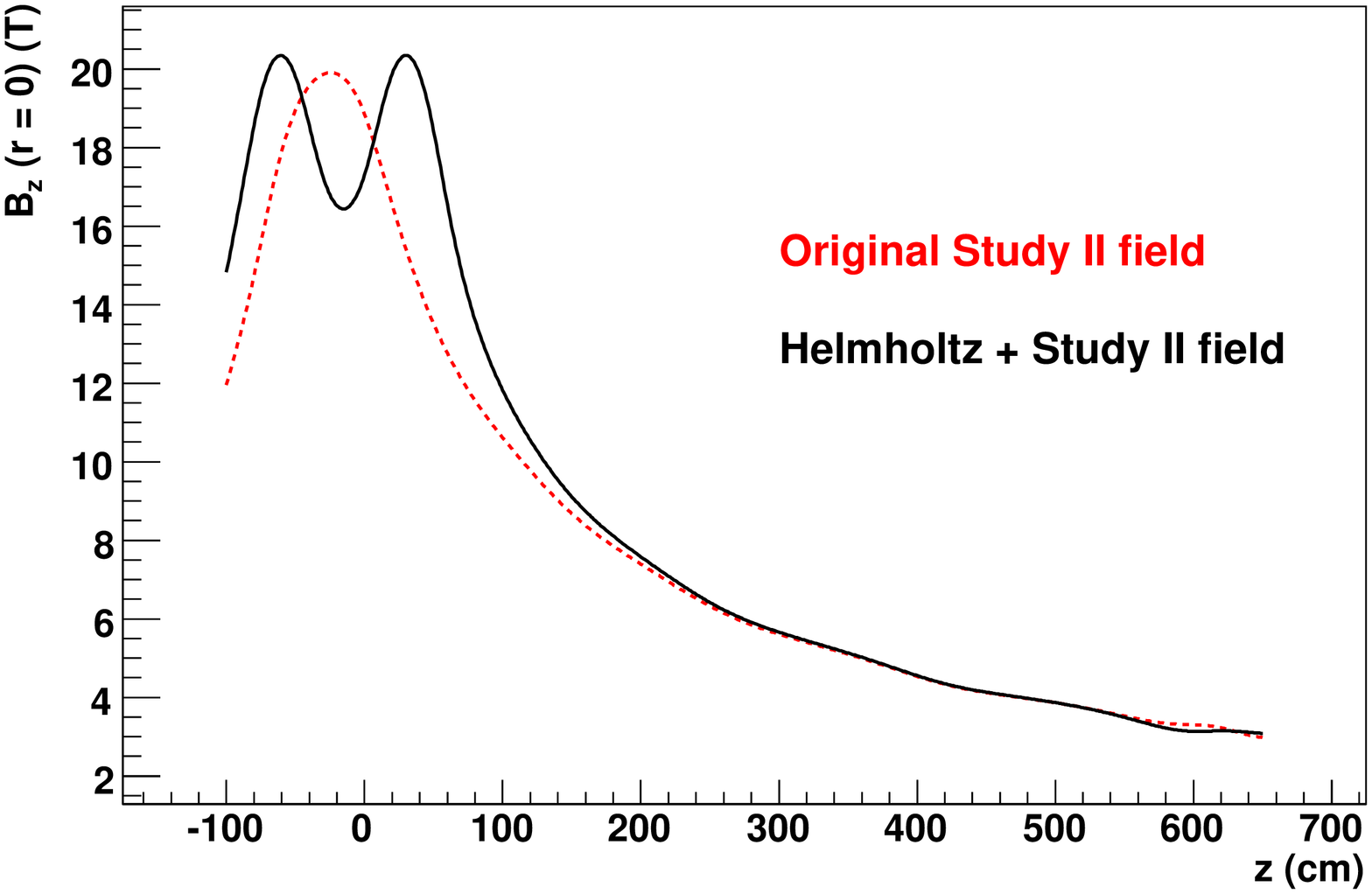, clip=, angle=-90, width=\textwidth}
\end{minipage}\hfill
\caption{Left: Schematic geometry ($x$ vs $z$) of the Helmholtz solid
target arrangement. The colour scheme is: target cylindrical rod (black), 
magnetic field lines (red arrows), copper coils (magenta), 
superconducting magnets (yellow), tungsten-carbide shielding (brown),
iron plug (purple). Right: Comparison of the $z$-component of the magnetic 
field along the solenoidal $z$ axis between the Helmholtz (solid line) and
Study-II~\cite{Study2} (dotted red line) geometries.}
\label{fig:geometry}
\end{center}
\end{figure}

The baseline Neutrino Factory target geometry is the Study-II~\cite{Study2} design
for a mercury jet system. The same geometry can be used for a powdered ($\leq 50\%$ density)
tungsten jet. However, some changes are needed for a solid target system. 
One option is to equally space cylindrical tungsten rods along the circumference of a spokeless 
wheel~\cite{RobsPaper}. To achieve this, a gap needs to be introduced inside the solenoids 
(analogous to a Helmholtz coil) so that the wheel can pass through the target aperture.
Figure~\ref{fig:geometry} shows a schematic of the $x-z$ geometry.
The gap is just long enough to ensure the solid target rod can pass through.
Note that shielding will be required along the gap to stop radiation
reaching the superconducting coils. A first estimate of the shielding requirement 
is 10\,cm on either side of the gap, although this needs further study. Additionally, 
the average current density inside the copper coils needs to be increased from 
20\,A\,mm$^{-2}$ (Study-II) to 30\,A\,mm$^{-2}$ to achieve a peak magnetic field 
of 20\,T inside the target aperture.

\section{Pion and muon yields}

\begin{figure}[hbt!]
\begin{center}
\epsfig{file=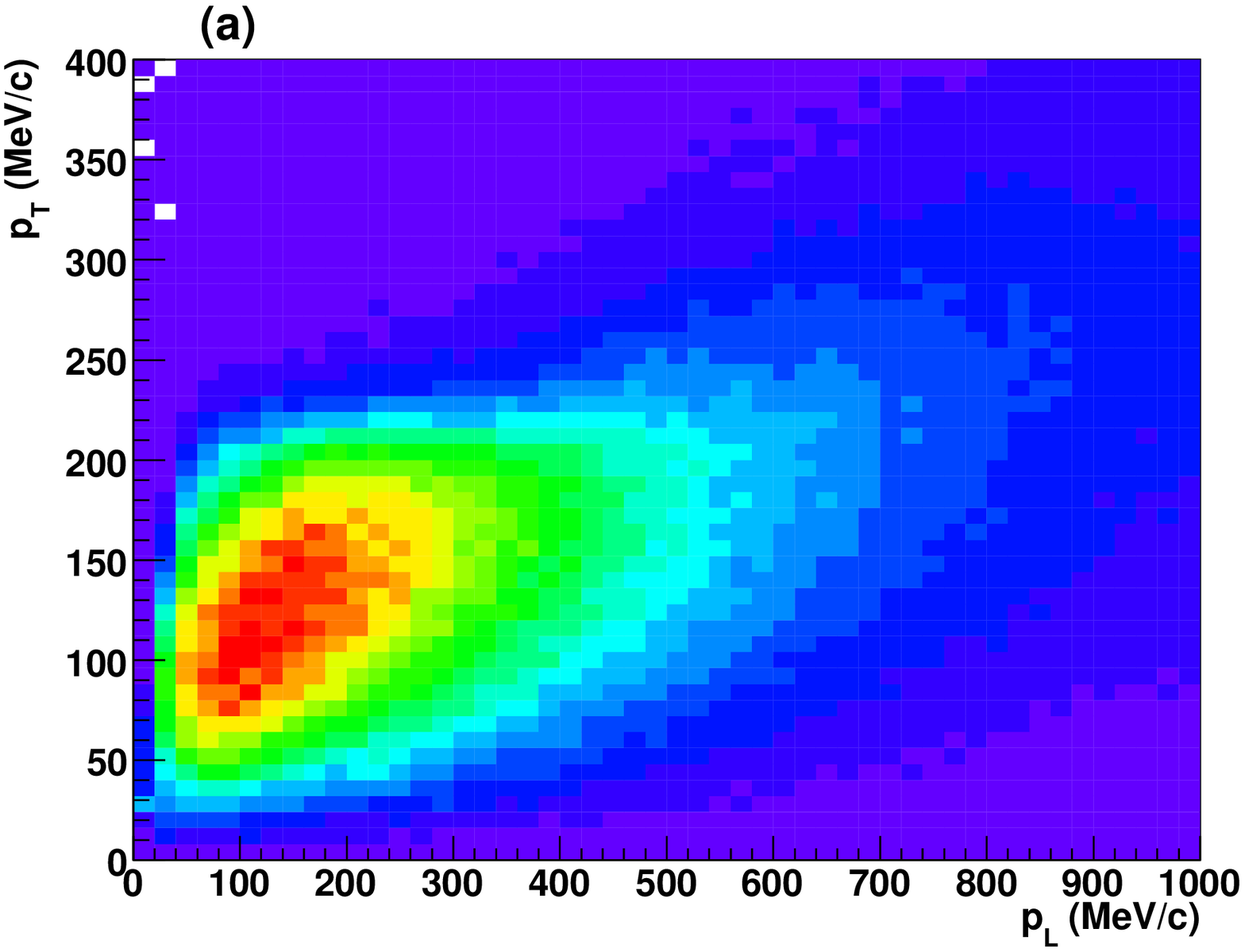, clip=, width=0.45\textwidth}
\epsfig{file=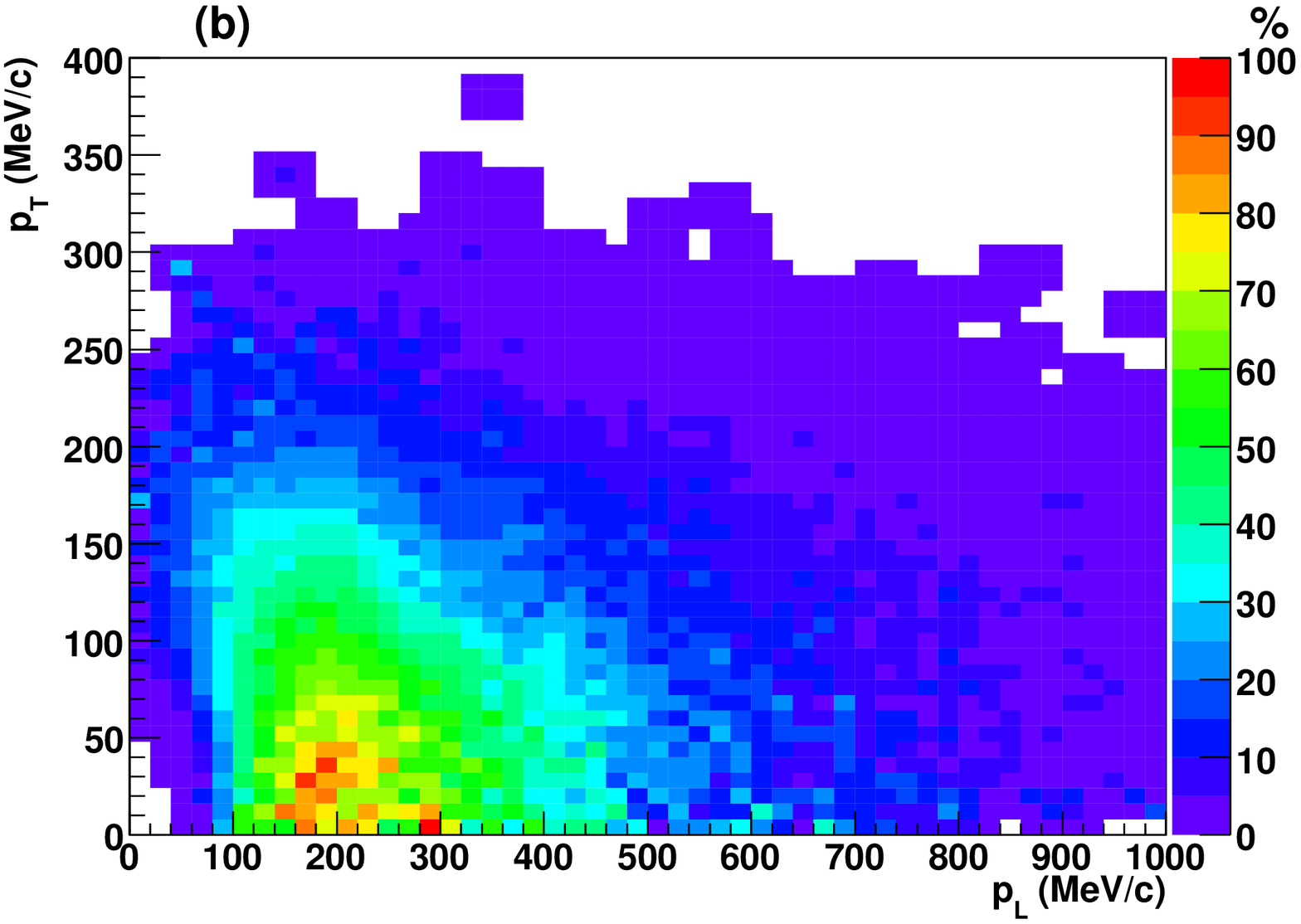, clip=, width=0.45\textwidth}
\caption{(a) Density-plot of the initial longitudinal ($p_L$) and 
transverse ($p_T$) momenta of pions at $z = 0$\,cm. (b) Probability acceptance
map of pions from the target going through the Neutrino Factory cooling channel 
as a function of the initial longitudinal and transverse momenta.}
\label{fig:acceptance}
\end{center}
\end{figure}

Here, we present the pion and muon yields from MARS~\cite{MARS} simulations of a 
parabolic 10\,GeV proton beam interacting with solid or powdered (50\% density) 
tungsten targets. The number of pions and muons produced at the 
target are corrected for the acceptance through the whole muon cooling channel. 
Figure~\ref{fig:acceptance}a shows an example density-plot of the initial longitudinal 
$(p_L)$ and transverse $(p_T)$ momenta of pions at the end of the target 
rod inside the solenoid aperture ($z = 0$\,cm, $r \leq 7.5$\,cm).
Figure~\ref{fig:acceptance}b shows the probability of pions (and muons) from
the target reaching the end of the muon cooling channel, which was obtained
by passing the input momentum distribution shown in Fig.~\ref{fig:acceptance}a 
through an ICOOL~\cite{ICOOL} simulation of the Neutrino Factory muon cooling channel.

\begin{figure}[htb!]
\begin{center}
\epsfig{file=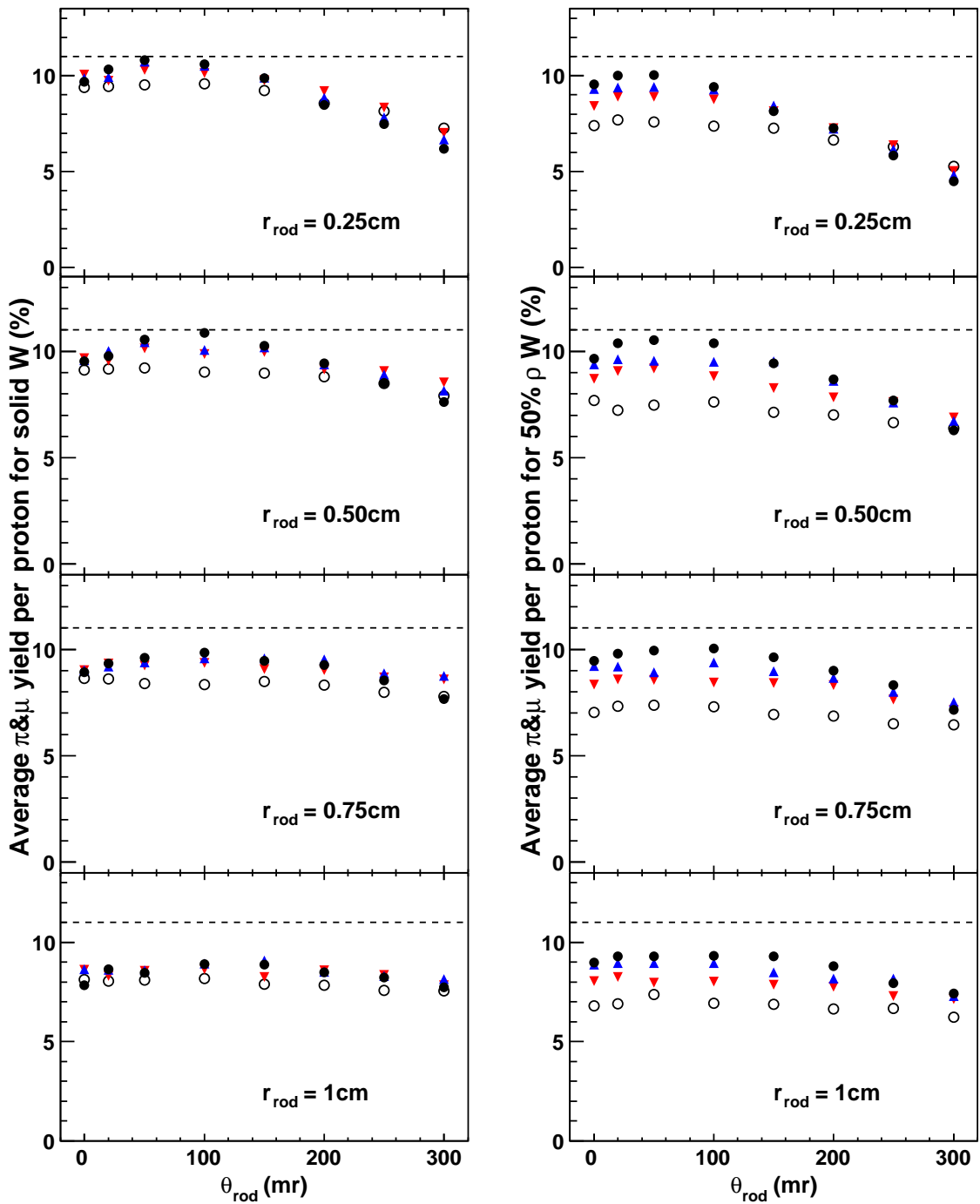, clip=, angle=0, width=\textwidth}
\caption{Charge-averaged accepted $\pi$ and $\mu$ yield per 10\,GeV proton (\%) as a 
function of target rod tilt ($\theta_{\rm{beam}} = \theta_{\rm{rod}}$) for the solid 
tungsten (Helmholtz geometry) and 50\% density tungsten (Study-II geometry) targets. 
Each row corresponds to a specific beam and target radius ($r_{\rm{beam}} = r_{\rm{rod}}$),
while each point corresponds to a different target rod length: 15\,cm (open circle), 
20\,cm (lower red triangle), 25\,cm (upper blue triangle) and 30\,cm (solid circle).
The dotted horizontal line represents the yield for the Neutrino Factory mercury jet target
($\theta_{\rm{beam}} = 67$\,mr, $\theta_{\rm{Hg}} = 100$\,mr,
$r_{\rm{beam}} = 0.15$\,cm, $r_{\rm{Hg}} = 0.50$\,cm).}
\label{fig:yields}
\end{center}
\end{figure}

Figure~\ref{fig:yields} shows the charge-averaged accepted pion and muon
yields per proton for solid cylindrical tungsten rods, with various
lengths and radii, as a function of rod tilt ($\theta_{\rm{rod}}$) with respect to the $z$ axis.
Also shown are the yields for the 50\% density (powdered) jet, which
is modelled as a simple cylinder (analogous to the solid material case),
whose length represents the effective length of the proton beam intersecting the jet.
The parabolic proton beam has a radius ($r_{\rm{beam}}$) and tilt ($\theta_{\rm{beam}}$)
equal to that of the target rod for both target
materials. The dotted horizontal line represents the accepted pion and muon yield for the 
mercury jet target scenario ($\theta_{\rm{beam}} = 67$\,mr, $\theta_{\rm{Hg}} = 100$\,mr,
$r_{\rm{beam}} = 0.15$\,cm, $r_{\rm{Hg}} = 0.50$\,cm).
The yields for the solid tungsten target are comparable (lower)
than those for the mercury jet when $r_{\rm{rod}} < 0.75$\,cm 
($r_{\rm{rod}} > 0.75$\,cm). The overall optimal target tilt is approximately 100\,mr 
and the optimal length of the target is $\geq 25$\,cm ($\geq 2.6$ interaction lengths).
It is interesting to note that, surprisingly, the yields for the 50\% density (powdered)
tungsten target are comparable to those for the solid target. 
One possibility for this is that, even though a smaller number of protons will interact with
the target owing to its lower density, there will be less re-absorption of pions 
inside the material. Secondly, a larger fraction of pions will get focused 
inside the solenoid aperture owing to a stronger ($\sim$20\%) magnetic field at the end of
the target rod ($z = 0$\,cm). Further work is ongoing to compare these results with 
Geant4~\cite{Geant4} simulations.

\section{Pion re-absorption}

The size and position of the target must be carefully chosen to ensure
that an adequate number of pions are produced from proton-nucleon interactions 
with a minimal number of pions getting re-absorbed inside the material. Table~\ref{tab:absorb}
shows a comparison of the fractional loss $f_{\rm{lost}}$ of the charge-averaged pion
yield $Y$ for various solid target scenarios compared to the yield $Y_0$ 
from one 30\,cm-long rod ($\theta_{\rm{rod}} = \theta_1$): $f_{\rm{lost}} = (Y_0 - Y)/Y_0$. 
The first scenario uses three 30\,cm-long target rods separated by 10\,cm along the $z$ axis, 
the second uses a horizontal solid target toroid with a radius of curvature of 5\,m 
(effective length of 1\,m along $z$) and tube radius $r_{\rm{rod}}$, 
while the third is similar to the three rod case, 
but with all targets tilted at 100\,mr with respect to the $z$ axis ($\theta_{\rm{rods}} = \theta_3$).
The 10\,GeV proton beam is tilted at $\theta_{\rm{beam}} = 67$\,mr for all cases.
One can see that there is significant pion absorption
for the first two cases; tilting the target rods dramatically reduces this effect.
\begin{table}[!hb]
\begin{center}
\begin{tabular}{|c|c|c|c|c|}
\hline
$r_{\rm{rod}}$ & $r_{\rm{beam}}$ & 3 rods  $f_{\rm{lost}}$ & toroid  $f_{\rm{lost}}$ & 3 rods  $f_{\rm{lost}}$ \\
& & $(\theta_1 = \theta_3 = 0)$ & ($\theta_1 = 0$) & ($\theta_1 = \theta_3 = 100$\,mr) \\
\hline
0.5\,cm & 0.5\,cm & 39\% & 23\% & 7\% \\
1\,cm   & 1\,cm   & 57\% & 47\% & 15\% \\
1\,cm   & 1.5\,cm & 53\% & 53\% & 11\% \\
\hline
\end{tabular}
\caption{The fractional loss of charge-averaged pion yields for different solid target scenarios.}
\label{tab:absorb}
\end{center}
\end{table}

The amount of pion re-absorption for the mercury jet target has also been estimated.
Consider the target aperture up to the location of the beryllium window at $z=2$\,m
to be filled with mercury vapour with an effective density given
by $\rho_{\rm{vapour}} = \rho_{\rm{He}}\times({\rm{0.1\,bar}/1\,atm}) + w \rho_{\rm{Hg}}$,
where $\rho_{\rm{He}}$ is the density of the He gas and $w$ is the ratio of the volume of the mercury jet
to the total volume inside the solenoid aperture (0.3\%). In this worse case scenario, the 
relative fractional loss of pions by re-absorption in the vapour is 5\%.

\section{Conclusion}

We have shown results from MARS simulations of the accepted pion and muon
yields when a 10\,GeV parabolic proton beam interacts with a solid or powdered 
(50\% density) tungsten target for a Neutrino Factory. These yields are comparable
to those for the mercury jet target. We have also provided estimates
of pion re-absorption for solid tungsten and mercury jet targets.

\end{document}